# Cluster formation probability in the trans-tin and trans-lead nuclei

K P Santhosh*, R K Biju and Sabina Sahadevan

School of Pure and Applied Physics, Kannur University, Payyanur Campus, Payyanur 670 327, India

**Abstract**

Within our fission model, the Coulomb and proximity potential model (CPPM) cluster formation probabilities are calculated for different clusters ranging from carbon to silicon for the parents in the trans-tin and trans- lead regions. It is found that in trans-tin region the $^{12}$C, $^{16}$O, $^{20}$Ne and $^{24}$Mg clusters have maximum cluster formation probability and lowest half lives as compared to other clusters. In trans-lead region the $^{14}$C, $^{18,20}$O, $^{23}$F, $^{24,26}$Ne, $^{28,30}$Mg and $^{34}$Si clusters have the maximum cluster formation probability and minimum half life, which show that alpha like clusters are most probable for emission from trans-tin region while non-alpha clusters are probable from trans-lead region. These results stress the role of neutron proton symmetry and asymmetry of daughter nuclei in these two cases.

* drkpsanthosh@gmail.com

PACS number(s): 23.70.+j ; 23.60.+e ; 21.60.Gx
*Keywords*: Cluster radioactivity, Alpha decay, Clustering in nuclei

## 1. Introduction

Cluster radioactivity which is the spontaneous emission of particles heavier than alpha particle can be treated as an intermediate process between alpha decay and spontaneous fission, This phenomenon was first predicted by Sandulescu et al. [1] in 1980 on the basis of quantum mechanical fragmentation theory (QMFT) [2]. After four years Rose and Jones [3] confirmed this phenomenon by the emission of $^{14}$C emission from $^{223}$Ra. Generally there are two approaches in explaining the cluster decay process: (1) cluster model and (2) fission model. In cluster model [4–6] the cluster is assumed to be preformed in the parent nuclei before it penetrates the nuclear interacting barrier. In fission model [7–9] the nucleus deforms continuously as it penetrates the nuclear barrier and reaches the scission configuration after running down the Coulomb barrier. In the cluster decay studies cluster-like shapes are preferred for very high asymmetric mass splitting and fissioning shapes are most suitable for less asymmetric and symmetric mass splitting.

Clustering is a very general phenomenon, which appears in atomic, nuclear, sub nuclear, and the cosmic worlds [10]. A direct indication of clustering is the emission of alpha particle or heavier clusters from heavy nuclei. The properties of light nuclei with cluster structures have been well studied with cluster model. Clustering properties of nuclei would be helpful to obtain a systematic understanding of both the stable and exotic nuclei. Examples of such successful frameworks are the methods of fermionic molecular dynamics [11] and antisymmetrised molecular dynamics [12] which describe the structural properties of several nuclei, and their excited states, in the lighter mass region [13-16].

In nuclear dynamics, as seen in light stable nuclei, clustering is one of the essential features and various cluster structures have been known even in the low-energy region. Also, in the physics of unstable nuclei, clustering features comprise one of the central subjects. It is already well known [12, 13, 17-20] that clustering structures appear in the ground states of ordinary light nuclei with $N = Z$ or in their neighborhood. Even though this phenomenon has now been studied for a long time, many things remain to be learned and recently some experimental activities focused on this subject [21] were reported.

The present work aims to study the possibility of clustering in heavy nuclei by computing the cluster formation probability for the entire experimentally observed cluster decays from carbon to silicon taking the Coulomb and proximity potential as interacting barrier for the post-session region and simple power law interpolation for overlap region. We have also computed the cluster formation probability for different C, O, Ne and Mg clusters from $^{112,114}$Ba, $^{116,118}$Ce, $^{120,122}$Nd and $^{124,126}$Sm parents in the trans-tin region. The preformation probability can be calculated within a fission model as a penetrability of the internal part of the barrier, which corresponds to still overlapping fragments [22–24]. This part, however, is difficult to deal microscopically from current nuclear models due to the complexity of the nuclear many-body problem.

In our previous works, Santhosh et al. [25-30] proposed the Coulomb and proximity potential model to study cluster radioactivity in various proton-rich parents with Z = 56-64 and N = 56-72, leading to doubly magic$^{100}$Sn. We have studied the cold valleys in the radioactive decay of $^{248-254}$Cf isotopes [31] and the computed alpha decay half-lives closely agree with the experimental data. A semi-empirical model [32] is introduced for describing cluster radioactivity and this model has been applied to study the alpha decay of parents with $Z$ = 85–102. The competition between the alpha decay and spontaneous fission half-lives of even–even nuclei with $Z$ = 100–122 is studied [33] using the present model and the phenomenological formula for spontaneous fission. Recently we have proposed a semi empirical formula [34] for determining the spontaneous fission half lives which is applied to the mass region from $^{232}$Th to $^{286}$114. A brief description of cluster formation probability is given in section 2, the details of the Coulomb and proximity potential model are given in section 3 and the results and discussions are given in section 4.

## 2. Cluster formation Probability

The first model used to compute preformation probability was proposed by Blendowske, Flessbach and Walliser [4, 35, 36] on the basis of microscopic wave functions. For the case of $^{14}C$ emission from $^{223}Ra$, the authors proposed an empirical formula for computing the cluster formation probability and is given as,

$$S = (S_\alpha)^{(A_C - 1)/3} \tag{1}$$

here $S_\alpha$ is the α-particle preformation probability and the values of S are different for even and odd nuclei. For even–even parent $S_\alpha^{even} = 0.0063$ and for odd-A parent nuclei, $S_\alpha^{odd} = 0.0032$ [35]. Recently Dong et al. [37] computed the preformation probability using the eqn. 1 with the value of $S_\alpha = 0.0338$ for even-even parent and $S_\alpha = 0.0262$ for odd-A parents. We have computed the cluster formation probability for $^{14}C$ cluster emission from even-even parents by using the same relation and is found that $S_{\text{Blendowske}} = 2.91x10^{-10}, S_{Dong} = 4.22x10^{-7}$.

Gupta and co workers [5] considered that in nuclear decay the formation of two fragments (the cluster and daughter nucleus) in their ground state with probability $P_0$, defined the as a quantum mechanical probability of finding the fragments $A_1$ and $A_2$ (with fixed charges $Z_1$ and $Z_2$, respectively) at a point R of the relative motion, the cluster formation probability is given as

$$P_0(A_1) = \left|\psi_{R\eta_Z}^{(0)}(A_1)\right|^2 \sqrt{B_{\eta\eta}(A_1)}\,\frac{2}{A} \tag{2}$$

where $B_{\eta\eta}$ represents the mass parameter. The quantum number $\omega = 0,1,2,\ldots..$ counts the vibrational states $\psi_{R\eta_Z}^{(\omega)}$ in the potential V.

Within fission model Poenaru et al [38] defined the preformation probability, S as the penetrability through the internal part (overlap region) of the barrier:

$$S = \exp(-K_{ov}) \tag{3}$$

Where

$$K_{ov} = \frac{2}{\hbar}\int_{R_a}^{R_t} \sqrt{2B(R)[E(R)-Q)}\,dR \tag{4}$$

In which $B(R)$ is the nuclear inertia, $E(R)$ is the deformation energy and $R_a$ the internal turning point, defined by $E(R_a) = Q$, the energy released.

There are many empirical relations put forward for computing the cluster formation probability, $P_0$ based on different models. These are

$-\log_{10} P_0 = 0.363\,A_C + 3.044$ for $A_C < 30$ Malik – Gupta [5]

$-\log_{10} P_0 = 0.309\,A_C + 2.214$ Sandulescu et al [39] (5)

$-\log_{10} P_0 = 0.733\,A_C - 0.733$ for even A Blendowske – Walliser [4]

For $A_C < 30$ the three relations display the straight line description. For $A_C > 30$, only the calculations of Malik and Gupta are available for heavier cluster emissions and it is found that the cluster formation probability increases, rather than decreases, and becomes almost constant for $A_C > 40$. We would like to point out that in our earlier works [40] we have computed the cluster formation probability for all the clusters ranging from $^4$He to $^{24}$Ne for even-even parents with Z = 100-112 and found that $P_0$ decreases with increase in mass number of cluster up to $A_C = 20$ and then remains a constant. It is to be noted that different models [4, 5, 37-39] give different values for cluster formation probability but the computed $T_{1/2}$ values agrees with experimental data which imply that $P_0$ depends on models.

## 3. The Coulomb and Proximity Potential Model

The interacting potential barrier for a parent nucleus exhibiting cluster decay is given by

$$V = \frac{Z_1 Z_2 e^2}{r} + V_p(z) + \frac{\hbar^2 \ell(\ell+1)}{2\mu r^2} \qquad \text{, for } z > 0 \qquad (6)$$

Here $Z_1$ and $Z_2$ are the atomic numbers of daughter and emitted cluster, 'r' is the distance between fragment centers, 'z' is the distance between the near surfaces of the fragments, $\ell$ the angular momentum, $\mu$ the reduced mass and Vp is the proximity potential is given by Blocki et al. [41]

$$V_p(z) = 4\pi\gamma b \left[ \frac{C_1 C_2}{(C_1 + C_2)} \right] \Phi\left( \frac{z}{b} \right) \qquad (7)$$

With the nuclear surface tension coefficient,

$$\gamma = 0.9517\,[1 - 1.7826\,(N - Z)^2 / A^2] \qquad \text{MeV/fm}^2 \qquad (8)$$

Where N, Z and A represent neutron, proton and mass number of parent, $\Phi$, represent the universal the proximity potential is given as [42]

$$\Phi(\varepsilon) = -4.41 e^{-\varepsilon/0.7176} \qquad \text{for } \varepsilon \geq 1.9475 \qquad (9)$$

$$\Phi(\varepsilon) = -1.7817 + 0.9270\,\varepsilon + 0.0169\,\varepsilon^2 - 0.05148\,\varepsilon^3 \qquad \text{for } 0 \leq \varepsilon \leq 1.9475 \qquad (10)$$

With ε = z/b, where the width (diffuseness) of the nuclear surface b ≈1 and Siissmann central radii $C_i$ of fragments related to sharp radii $R_i$ is

$$C_i = R_i - \left( \frac{b^2}{R_i} \right) \qquad (11)$$

For $R_i$ we use semi empirical formula in terms of mass number $A_i$ as [41]

$$R_i = 1.28 A_i^{1/3} - 0.76 + 0.8 A_i^{-1/3} \qquad (12)$$

The potential for the internal part (overlap region) of the barrier is given as

$$V = a_0 (L - L_0)^n \qquad \text{for } z < 0 \qquad (13)$$

here $L = z + 2C_1 + 2C_2$ and $L_0 = 2C$, the diameter of the parent nuclei. The constants $a_0$ and $n$ are determined by the smooth matching of the two potentials at the touching point.

Using one dimensional WKB approximation, the barrier penetrability P is given as

$$\mathrm{P} = \exp\left\{ -\frac{2}{\hbar} \int_a^b \sqrt{2\mu (V - Q)}\, dz \right\} \qquad (14)$$

Here the mass parameter is replaced by reduced mass $\mu = m A_1 A_2 / A$, where m is the nucleon mass and $A_1$, $A_2$ are the mass numbers of daughter and emitted cluster respectively. The turning points "a" and "b" are determined from the equation $V(a) = V(b) = Q$. The above integral can be evaluated numerically or analytically, and the half life time is given by

$$T_{1/2} = \left( \frac{\ln 2}{\lambda} \right) = \left( \frac{\ln 2}{\upsilon P} \right) \qquad (15)$$

Where, $\upsilon = \left( \frac{\omega}{2\pi} \right) = \left( \frac{2E_v}{h} \right)$ represent the number of assaults on the barrier per second and λ the decay constant. $E_v$, the empirical zero point vibration energy is given as [8]

$$E_v = Q \left\{ 0.056 + 0.039 \exp\left[ \frac{(4 - A_2)}{2.5} \right] \right\} \qquad \text{for } A_2 \geq 4 \qquad (16)$$

For alpha decay, $A_2 = 4$ and the empirical zero point vibration energy becomes, $E_v = 0.095 Q$

The cluster formation probability, S can be calculated within the fission model as the penetrability of the internal part (overlap region) of the barrier given as

$$S = \exp(-K) \tag{17}$$

Where

$$K = \frac{2}{\hbar}\int_{a}^{0} \sqrt{2\mu(V-Q)}\,dz \tag{18}$$

Here, $a$ is the inner turning point and is defined as $V(a) = Q$ and z = 0 represents the touching configuration.

## 4. Results and Discussion

We have computed the cluster formation probability for the clusters ranging from carbon to silicon from the parent nuclei in trans-tin and trans-lead region using eqn. 18. Fig. 1 represents the plot for computed $\log_{10}$ (S) versus $A_2$, mass number of the fragment, for the emission of different carbon clusters from $^{112,114}$Ba, oxygen clusters from $^{116,118}$Ce, neon clusters from $^{120,122}$Nd and magnesium clusters from $^{124,126}$Sm parents respectively. The Q values are computed using the experimental binding energies of Audi et al [43] and some masses are taken from the table of KTUY [44]. It is clear from the plot that the maximum value for cluster formation probability is obtained for $^{12}$C, $^{16}$O, $^{20}$Ne and $^{24}$Mg clusters from the corresponding parents. One of the interesting facts is that the computed Q-values are high for these decays as compared to other reactions. As considered the neutron proton asymmetry distribution in light clusters by the $\frac{N}{Z}$ ratio of the matter at the centre of the nuclei, for N = Z matter means that the neutron and proton distribution is symmetrical and only the alpha particle like matter exist at the centre of the nuclei. These nuclei are found to be in the neutron proton ratio equal to 1 and having greater cluster formation probability from the parent nucleus. Therefore we would like to point out that the clusters $^{12}C$, $^{16}O$, $^{20}Ne$ and $^{24}Mg$ form alpha like cluster configurations in the parent nucleus.

Arumugam et al. [45] studied the possibility of clustering in light clusters with the increase or decrease of neutron numbers using relativistic mean field theory and examined the internal structure (density distribution) of various Be, C, O, Ne, Mg and Si isotopes. The authors found that the density distribution of $^{6}Be$ indicates a $\alpha + 2p$ structure since there is more configuration of nuclear matter in the central region, the $^{8}Be$ nuclei can easily be thought of as an α-α cluster configuration. As the neutron concentration increases for Be nuclei, heavier than $^{8}Be$, the clustering in proton matter remains almost undisturbed. i.e., the core having alpha-alpha cluster persists with the addition of neutron to the alpha-alpha nucleus $^{8}Be$, i.e. these isotopes have an $\alpha + \alpha + xn$ structure, with alpha-alpha as the core. The remaining $x$ neutrons are rather sparsely distributed around the core and constitutes a halo for these heavier isotopes of Be. For the case of carbon isotopes Freer et al. [18] and Horiuchi et al. [19] predicted the 3α-chain structure of $^{12}C$ isotope. In the case of oxygen isotopes the $^{16}O$ is a doubly magic nucleus and it is spherical in shape. Zhang et al. [46] studied the structure of $^{16}O$ isotope and found that it forms a square, rhombic or a kite structure with alpha nucleus at its corner. The RMF calculations predict the ground state of $^{16}O$ as having four alpha particles in the form of a regular tetrahedron in three-dimensional configuration and in two-dimensional configuration they form a kite structure, (see Fig. 6(b) of Ref. [45]) at deformation β = 0.95 and it yields the linear $4\alpha$-chain structure at highest deformation (β = 3.79). In 1994 Zhang et al [47] predicted ground state of $^{20}Ne$ to be 5α-trigonal bipyramid and later the same structure was confirmed in many other calculations [4, 11]. Most of the calculations [47, 48] predict the ground state of $^{24}Mg$ as a $\alpha + {}^{16}O + \alpha$ structure with β~0.6 and having triaxial shape [49]. In relativistic mean field mechanism the $^{24}$Mg isotope has the $^{12}$C+ $^{12}$C structure, though with different orientations - prolate (central bispheroid) and oblate (trigonal biprism) and each $^{12}$C isotope is the combination of three alpha particles. Many authors [16, 20, 47] predicts that the $^{28}$Si isotopes have the ${}^{12}C + \alpha + {}^{12}C$ trigonal biprism shape. From these observations we can see that the heavy clusters (from Be to Si) are formed as the combination of alpha particles and additional neutrons ( $n\alpha + xn$ ).

Fig. 2 represents the plot for $\log_{10}$ (S) versus $A_{2,}$ mass number of various carbon clusters from Ra, Fr, Ac and Th parents. It is evident from the plot that a peak in cluster formation probability is obtained at $^{14}$C nucleus. We would like to point out that the $^{14}$C cluster emission from different Ra, Fr, Ac and Th parent nuclei has already been detected experimentally [50-55]. So the $^{14}$C cluster preforms in these parents as compared to other carbon clusters, we would like to point out that the $^{14}$C cluster form $(3\alpha + 2n)$ structure.

In Fig. 2 we have found a second peak at $^{16}$C cluster but the $^{16}$C cluster emissions are not observed experimentally. We have computed the half lives for $^{14,\ 16}$C cluster emissions from $^{224,\ 226}$Ra and $^{225}$Ac parents using Coulomb and proximity potential model. It is found that the lowest half lives is obtained for the $^{14}$C cluster emissions from these parents. {For e.g. in the case of $^{224}$Ra, $\log_{10}(T_{1/2}) = 16.74$ (for $^{14}$C) and $\log_{10}(T_{1/2}) = 27.45$ (for $^{16}$C); in $^{226}$Ra, $\log_{10}(T_{1/2}) = 22.55$ (for $^{14}$C) and $\log_{10}(T_{1/2}) = 34.44$ (for $^{16}$C); in $^{225}$Ac, $\log_{10}(T_{1/2}) = 18.03$ (for $^{14}$C) and $\log_{10}(T_{1/2}) = 30.92$ (for $^{16}$C)} i.e. the $^{14}$C cluster is most probable for emission, which preforms faster in these parents as compared to the $^{16}$C cluster. We would like to point out that many authors [56-58] predicts $^{14}$C cluster emissions from these parents and our values are in agreement with their predictions.

The preformation probability of different oxygen isotopes from $^{226,228}Th$ and $^{21-25}F$ clusters from $^{231}Pa$ are shown in figure 3. It is obvious from the plot that in the case of $^{228}Th$ parent the maximum value of cluster formation probability is obtained for $^{20}O$ and in the case of $^{231}Pa$ isotope the maximum value is obtained for $^{23}F$ cluster. i.e., the cluster $^{20}O$ forms $(5\alpha + 4n)$ structures in the $^{228}Th$ parent nucleus and $^{23}F$ forms $(4\alpha + 1p + 6n)$ structures in $^{231}Pa$ isotope. It is obvious from the Figs. 2-3 that the cluster formation probability is greatest for $^{14}C$, $^{18}O$ and $^{20}O$ clusters from $^{226}Th$ parent. The experimentally observed cluster decay for $^{14}C$ and $^{18}O$ clusters has the same half life (experimental lower limit for $\log_{10} T_{1/2}^{Expt}(s) = 15.3$ [55]). As we compare the preformation probability for these isotopes it is found that the cluster formation probability is maximum for $^{14}C$ cluster (for $^{14}C$, $S = 1.98\times10^{-4}$; for $^{18}O, S = 2.27\times10^{-6}$ and for $^{20}O, S = 2.98\times10^{-6}$) . We have also computed the half lives for these cluster emissions using Coulomb and proximity potential model [32] and is found that the lowest half life time is for $^{14}C$ cluster emission $(\log_{10} T_{1/2}(s) = 18.74)$ as compared to $^{18}O$ $(\log_{10} T_{1/2}(s) = 19.29)$ and $^{20}O$ cluster emissions $(\log_{10} T_{1/2}(s) = 25.26)$. i.e., the $^{14}C$ cluster preforms faster in $^{226}Th$ parent as compared to other clusters.

Fig. 4 represents the plot for $\log_{10}(S)$ versus $A_2$ for the emission of different neon clusters from various U, Th and Pa parents. It is clear from the plot that two peaks are present at $A_2 = 24$ and 26. In the case of $^{232}Th$ and $^{234,236}U$ parents the $^{24,26}Ne$ clusters are experimentally detected [59-62] but $^{26}Ne$ has the maximum preformation probability compared to $^{24}Ne$ cluster. Again the parents $^{230}Th$, $^{231}Pa$, $^{232,233,235}U$ have the maximum cluster formation probability for $^{24}Ne$ cluster as compared to $^{26}Ne$ cluster, i.e these clusters are formed very fast in these parents as compared to other clusters. The $^{24}Ne$ cluster has the $(5\alpha + 4n)$ structure and $^{26}Ne$ cluster has the $(5\alpha + 6n)$ structure.

For the case of $^{230}$U parent, the $^{22}$Ne and $^{24}$Ne cluster decays are experimentally observed with same half life (experimental lower limit for $\log_{10} T_{1/2}(s) = 18.2$ [55]). It is evident from the Fig. 4 that the $^{24}$Ne cluster has the greater cluster formation probability. We have computed the half lives for these decays using Coulomb and proximity potential model [32]; we found that the lower half life time is for $^{24}$Ne cluster emission $(\log_{10} T_{1/2}(s) = 22.37)$ as compared to the $^{22}$Ne cluster emission $(\log_{10} T_{1/2}(s) = 22.60)$. We would like to point out that the $^{24}$Ne cluster is formed faster than the other clusters from these parents and it has a $(5\alpha + 4n)$ combination.

Fig. 5 represents the plot for $\log_{10}(S)$ versus $A_2$ for the emission of $^{26\text{-}32}$Mg clusters from various $^{232\text{-}236}$U, $^{237}$Np, and $^{236,238}$Pu parents. It is clear form the plot that the highest peaks are found at $^{28,30}$Mg, the experimentally observed cluster radioactivity. In the case of $^{232\text{-}235}$U, $^{236}$Pu the maximum preformation probability is obtained for $^{28}$Mg cluster. For $^{236}$U parent $^{28,30}$Mg clusters have same half life ($\log_{10} T_{1/2}^{Expt}(s) = 27.58$ [63]) and in the case of $^{238}$Pu parent $^{28,30}$Mg clusters have same half life ($\log_{10} T_{1/2}^{Expt}(s) = 25.70$ [64]) but the highest peaks in cluster formation probability is obtained for $^{30}$Mg as compared to $^{28}$Mg cluster. i.e. the $^{30}$Mg cluster has the greater chances for emission from these parents. In $^{237}$Np isotope the $^{30}$Mg cluster has the highest cluster formation probability. The $^{28}$Mg cluster forms $(6\alpha + 4n)$ combination while $^{30}$Mg cluster forms $(6\alpha + 6n)$ structure.

In the case of $^{232,233}$U parents the $^{24}$Ne and $^{28}$Mg cluster decays are experimentally observed and it is clear that the lowest half life is obtained for $^{24}$Ne cluster (for $^{232}$U, $\log_{10} T_{1/2}^{Expt}(s) = 21.08$ [65] and for $^{233}$U, $\log_{10} T_{1/2}^{Expt}(s) = 24.83$ [66]). We would like to point out that our computation [32] also gives the same results (for $^{232}$U, $\log_{10} T_{1/2}(s) = 20.72$ and for $^{233}$U, $\log_{10} T_{1/2}(s) = 24.15$), i.e. the $^{24}$Ne cluster is preformed faster in these parents as compared to other clusters. For the case of $^{234}$U parent the cluster radioactivity is observed for $^{24,26}$Ne cluster with same half life $(\log_{10} T_{1/2}^{Expt}(s) = 25.9)$ [60, 61] and for $^{28}$Mg cluster with $\log_{10} T_{1/2}^{Expt}(s) = 27.54$ [62], but the cluster formation probability is maximum for $^{26}$Ne cluster compared to other decays. We have computed the half life for these cluster emissions and it is found that the lowest half life time is for $^{26}$Ne cluster emission $(\log_{10} T_{1/2}(s) = 25.88)$ as compared to $^{24}$Ne cluster decay $(\log_{10} T_{1/2}(s) = 27.39)$. In the case of $^{236}$U isotope the $^{26}$Ne $(\log_{10} T_{1/2}^{Expt}(s) = 25.9)$ [62] and $^{28,30}$Mg $(\log_{10} T_{1/2}^{Expt}(s) = 27.58)$ [63] cluster decays are experimentally observed, i.e the $^{26}$Ne cluster emission has the lower half life. As we compared the cluster formation probability of these clusters it is found that the $^{26}$Ne cluster has the larger preformation probability value. So we would like to point out that this cluster preforms first in the parent nucleus.

The plot for $\log_{10}$(S) versus $A_2$ for the emission of $^{32\text{-}36}$Si clusters from $^{240}$Pu, $^{241}$Am and $^{242}$Cm parents are shown in Fig. 6. It is obvious from the plot that the $^{34}$Si cluster has the maximum cluster formation probability in these parents and it has a $(7\alpha + 6n)$ structure.

In the present study the maximum cluster formation probability is obtained for those decays which are experimentally observed. In brief our study reveals that $^{14}$C, $^{18,20}$O, $^{23}$F, $^{24,26}$Ne, $^{28,30}$Mg and $^{34}$Si clusters have the maximum cluster formation probability in trans- lead region and $^{12}$C, $^{16}$O, $^{20}$Ne and $^{24}$Mg clusters have the maximum preformation probability in trans-tin region and are due to the double shell closure of $^{208}$Pb and $^{100}$Sn daughter respectively. Alpha like structures are probable for emission (maximum cluster formation probability) from trans-tin region and non-alpha like structures are probable for emission from trans-lead region which shows the role of neutron proton symmetry and asymmetry of daughter nuclei in these two cases.

**References**

[1] A.Sandulescu, D.N.Poenaru and W.Greiner, Sov.J.Part.Nucl. **11** (1980) 528

[2] J A Maruhn, W Greiner and W Scheid, in:R Bock (Ed) Heavy Ion Collisions, Amsterdam, North Holland, vol 2, 1980, p 399

[3] H. J Rose and G A Jones, Nature (London) **307 (**1984) 245

[4] R Blendowske, T Fliessbach and H Walliser, Nucl. Phys. **A 464** (1987) 75

[5] S S Malik and R K Gupta, Phys. Rev. **C 39** (1989) 1992

[6] B Buck and A C Merchant, Phys. Rev. **C 39** (1989) 2097

[7] Y J Shi and W J Swiatecki, Nucl. Phys. **A 438 (**1985) 450

[8] D N Poenaru, M Ivascu, A Sandulescu and W Greiner, Phys. Rev. **C 32** (1985) 572

[9] G Shanmugam and B Kamalaharan, Phys. Rev. **C 38** (1988) 1377

[10] W Greiner, Z. Phys. **A 349** (1994) 315

[11] H Feldmeier and J Schnack, Rev. Mod. Phys. **72** (2000) 655

[12] Kanada-En'Yo Y and H Horiuchi, Prog. Theor. Phys. Suppl. **142** (2001) 205

[13] Kanada-En'Yo Y, M Kimura and H Horiuchi, C. R. Phys. **4** (2003) 497

[14] T Neff and H Feldmeier 2003 Arxiv:nucl-th/0312130

[15] Kanada-En'Yo Y, H Horiuchi and A Ono, Phys. Rev. C **52** (1995) 628

[16] Kanada-En'Yo Y, Proceedings of the 8th International Conference on Clustering Aspects of Nuclear Structure and Dynamics, Nara, Japan, 2003

[17] K Ikeda, N Takigawa and H Horiuchi, Prog. Theor. Phys. Suppl. **E68** (1968) 464

[18] M Freer and A C Merchant, J. Phys. G: Nucl. Part. Phys. **23** (1997) 261

[19] H Horiuchi, K Ikeda and Y Suzuki, Prog. Theor. Phys. Suppl. **52** (1972) 89

[20] B R Fulton, Contemporary Phys. **40** (1999) 299

[21] C Beck, Int. J. Mod. Phys. **E 13** (2004) 9

[22] O A P Tavares, E L Medeiros and M L Terranova, J. Phys. G: Nucl. Part. Phys. **31** (2005) 129

[23] D N Poenaru, Y Nagame, R A Gherghescu and W Greiner, Phys. Rev. **C 65** (2002) 054308

[24] D N Poenaru and W Greiner, Phys. Scr. **44** (1991) 427

[25] K P Santhosh and Antony Joseph, Pramana- J.Phys **55** (2000) 375

[26] K P Santhosh and Antony Joseph, Proc. Int. Nat. Symp. Nucl. Phys. (India) **43B** (2000) 296

[27] K P Santhosh and Antony Joseph, Pramana- J.Phys **58** (2002) 611

[28] K P Santhosh and Antony Joseph, Ind. J. of Pure & App. Phys. **42** (2004) 806

[29] K P Santhosh and Antony Joseph, Pramana – J. Phys. **52** (2004) 957

[30] K P Santhosh, R K Biju, S Sabina and Antony Joseph, Phys. Scr. **77** (2008) 065201

[31] R K Biju, Sabina Sahadevan, K P Santhosh and Antony Joseph, Pramana – J. Phys. **70** **(**2008) 617

[32] K P Santhosh, R K Biju and Antony Joseph, J. Phys. G: Nucl. Part. Phys. **35** (2008) 085102

[33] K P Santhosh, R K Biju and Sabina Sahadevan, J. Phys. G: Nucl. Part. Phys. **36** (2009) 115101

[34] K P Santhosh, R K Biju and Sabina Sahadevan *Nucl. Phys.* **A** 832 (2010) 220

[35] R Blendowske and H Walliser, Phys. Rev. Lett. **61** (1988) 1930

[36] R Blendowske, T Fliessbach and H Walliser, Z Phys. A **339** (1991) 121

[37] J M Dong, H F Zhang, J Q Li and W Scheid, Eur. Phys. J. A 41(2009) 197

[38] D N Poenaru and W Greiner, Nuclear Decay Modes, Institute of Physics Publishing, Bristol, 1996, p.275

[39] A Sandulescu, R K Gupta, W Greiner, F Carstoiu and M Horoi, Int. J. Mod. Phys **E 1** (1992) 379

[40] K P Santhosh and Antony Joseph, Pramana. J. Phys. **59** (2002) 599

[41] J Blocki, J Randrup, W J Swiatecki and C F Tsang, Ann. Phys (N.Y) **105** (1977) 427

[42] J Blocki and W J Swiatecki, Ann. Phys (N.Y) **132** (1987) 53

[43] G Audi, A H Wapstra and C Thivault, Nucl. Phys. **A 729** (2003) 337

[44] H Koura, T Tachibana, M Uno and M Yamada, Prog. Theor. Phys. **113** (2005) 305

[45] P Arumugam, B K Sharma, S K Patra and R K Gupta, Phys. Rev. **C 71** (2005) 064308

[46] J Zhang and W D M Rae, Nucl. Phys. **A 564** (1993) 252

[47] J Zhang, W D M Rae and A C Merchant, Nucl. Phys. **A 575** (1994) 61

[48] S Marsh and W D M Rae, Phys. Lett. **B 180** (1986) 185

[49] G Leander and S E Larsson, Nucl. Phys. **A 239** (1975) 93

[50] R Bonetti, C Chiesa, A Guglielmetti, C Migliorino, P Monti, A L Pasinetti and H L Ravn, Nucl. Phys.A **576 (**1994) 21

[51] P B Price, J D Stevenson, S W Barwick and H L Ravn, *Phys*. Rev. Lett. **54 (**1985) 297

[52] E Hourani, L Rosier, G Berrier-Ronsin, E Elayi, A C Mueller, G Rappenecker, G Rotbard, G Renou, A Libe and L Stab, Phys. Rev. **C 44 (**1991) 1424

[53] R Bonetti, C Chiesa, A Guglielmetti, R Matheoud, C Migliorino, A L Pasinetti and H L Ravn, Nucl. Phys. **A 562 (**1993) 32

[54] E Hourani, M Hussonnois, L Stab, L Brillard, S Gales and J P Schapira, Phys. Lett. **B 160 (**1985) 375

[55] M Hussonnois, J F LeDu, L Brillard, J Dalmasso and G Ardisson, Phys. Rev. **C 43 (**1991) 2599

[56] H G de Carvalho, J B Martins and O A P Tavares, Phys. Rev. **C 34 (**1986) 2561

[57] Y Ronen, Phys. Rev. **C 44 (**1991) R594

[58] O A P Tavares, L A M Roberto and E L Medeiros, Phys. Scr. **76** (2007) 375

[59] R Bonetti, C Chiesa, A Guglielmetti, R Matheoud, G Poli, V L Mikheev and S P Tretyakova, Phys. Rev. **C 51 (**1995) 2530

[60] R Bonetti, C Chiesa, A Guglielmetti, C Migliorino, A Cesan, M Terrani and P B Price, Phys. Rev. **C 44** (1991) 888

[61] K J Moody, E K Hulet, S Wang and P B Price, Phys. Rev. **C 39** (1989) 2445

[62] S P Tretyakova, Yu S Zamyatnin, V N Kovantsev, Yu S Korotkin, V L Mikheev and G A Timofeev, Z. Phys. A **333** (1989) 349

[63] S P Tretyakova, V L Mikheev, V A Ponomarenko, A N Golovchenko, A A Oglobin and V A Shigin, JETP Lett. **59 (**1994) 394

[64] S Wang, D Snowden-Ifft, P B Price, K J Moody and E K Hulet, Phys. Rev. **C 39** (1989) 1647

[65] S W Barwick, P B Price and J D Stevenson, Phys. Rev. **C 31** (1985) 1984

[66] S P Tretyakova, A,Sandulescu V L Mikheev, D Hasegan, I A Lebedev, Yu S Zamyatnin, Yu S Korotkin and B F Myasoedov, JINR Dubna Rapid Commun. **13** (1985) 34

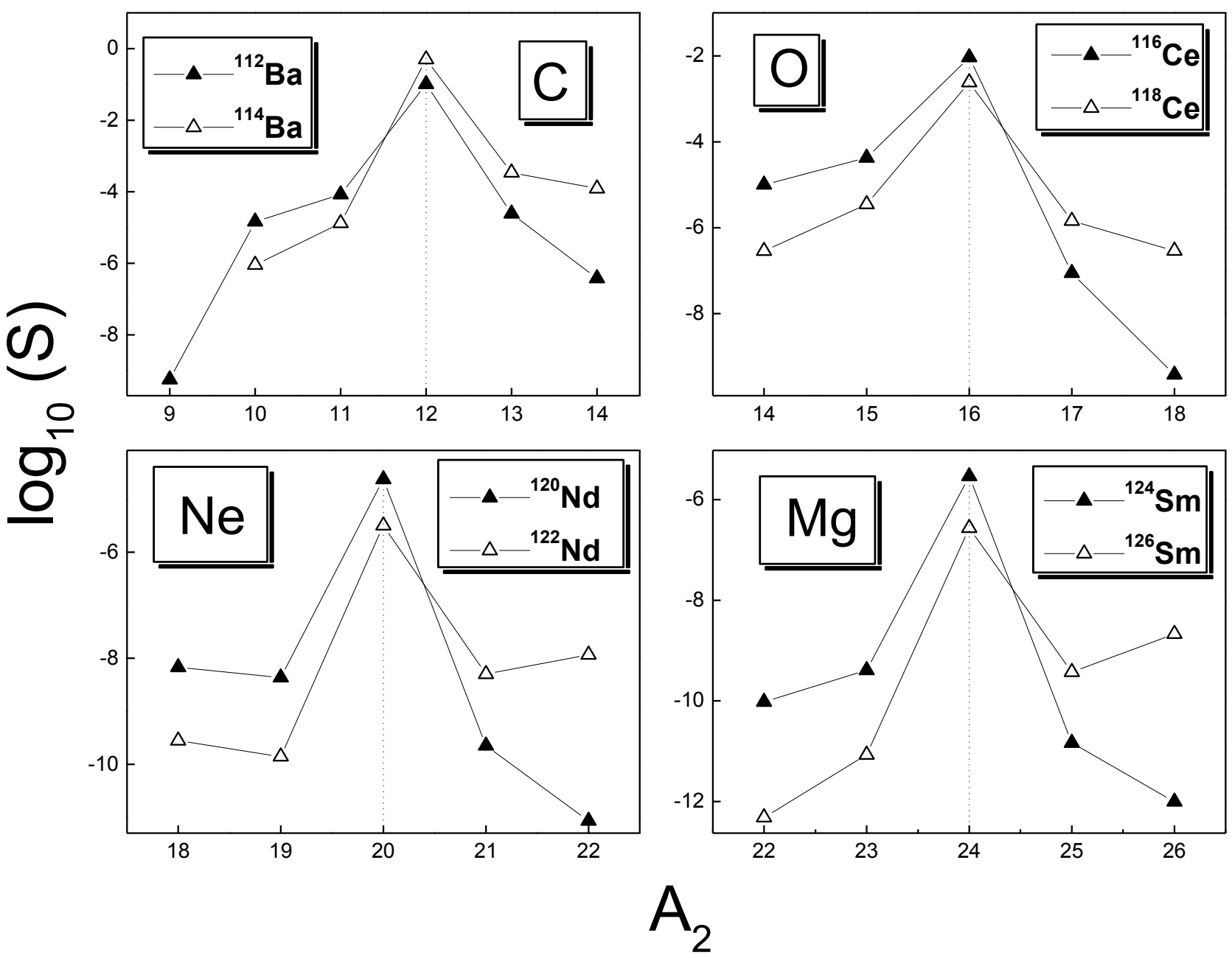

**Fig. 1.** Plot for $\log_{10}$ (S) versus $A_2$, the mass number of various C, O, Ne and Mg clusters from Ba, Ce, Nd and Sm parents.

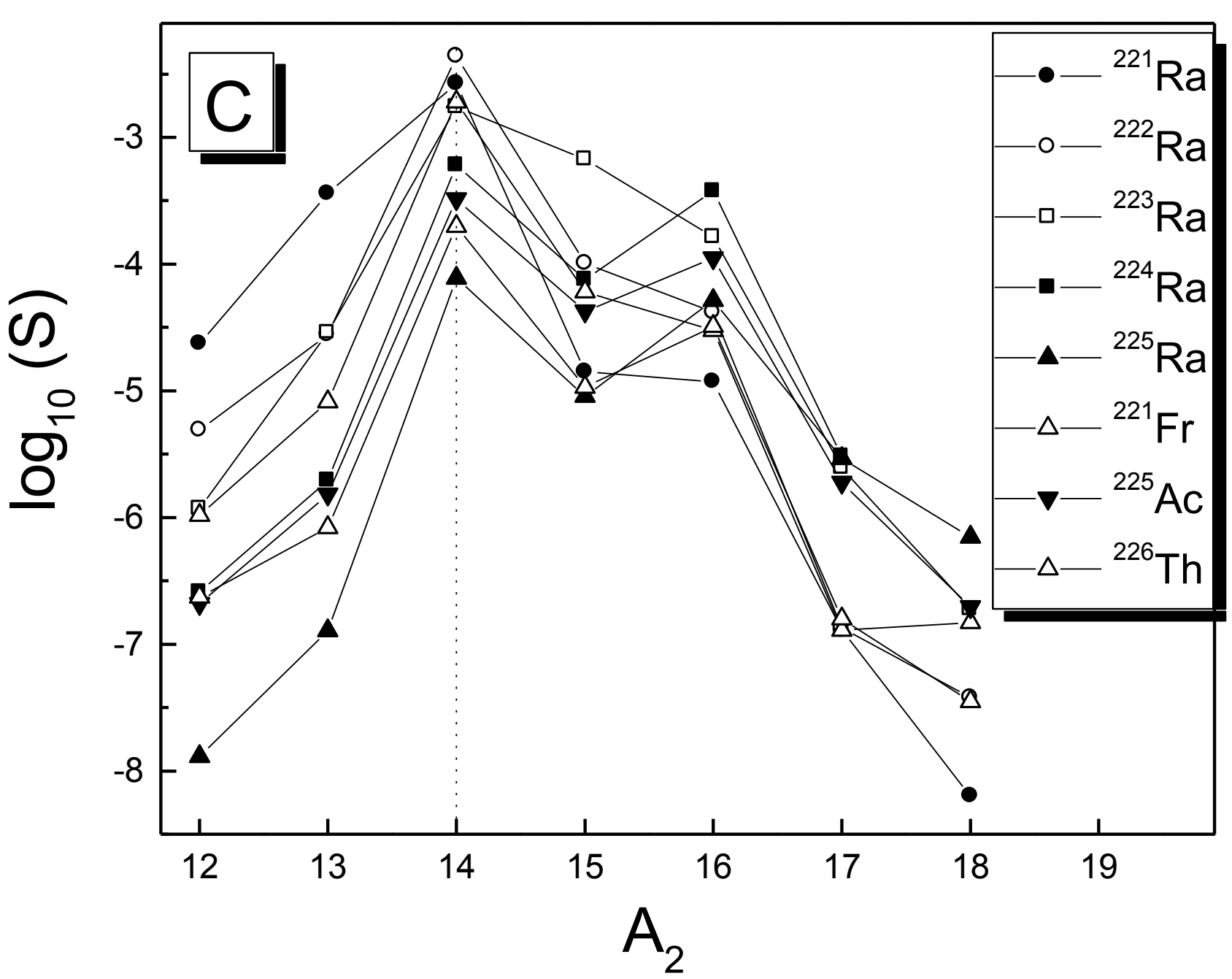

**Fig. 2.** Plot for $\log_{10}$ (S) versus $A_2$, the mass number of various carbon clusters from Ra, Fr, Ac and Th parents.

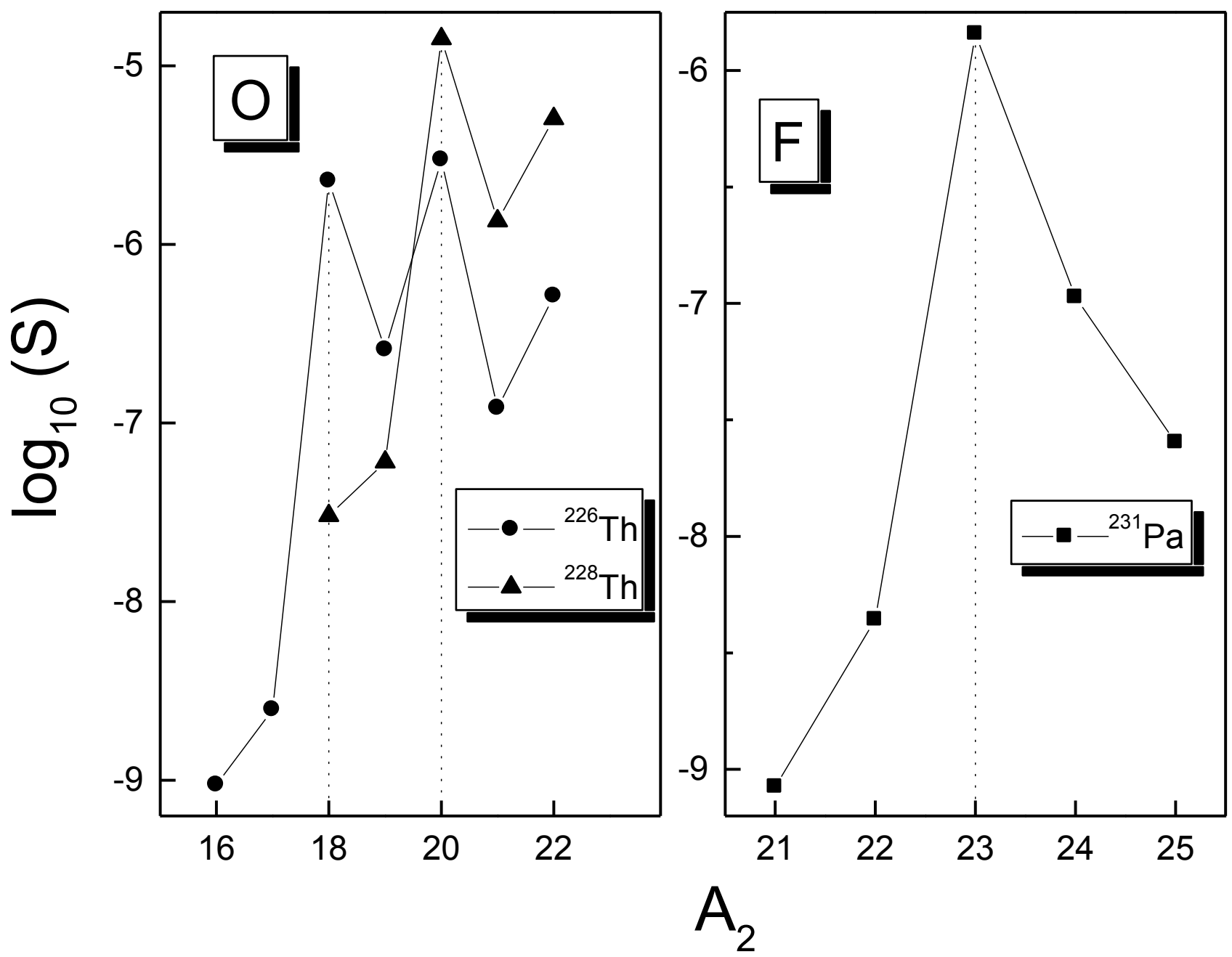

**Fig. 3.** Plot for $\log_{10}$ (S) versus $A_2$, the mass number of various oxygen clusters from $^{226,228}$Th and $^{21-25}$F clusters from $^{231}$Pa parents.

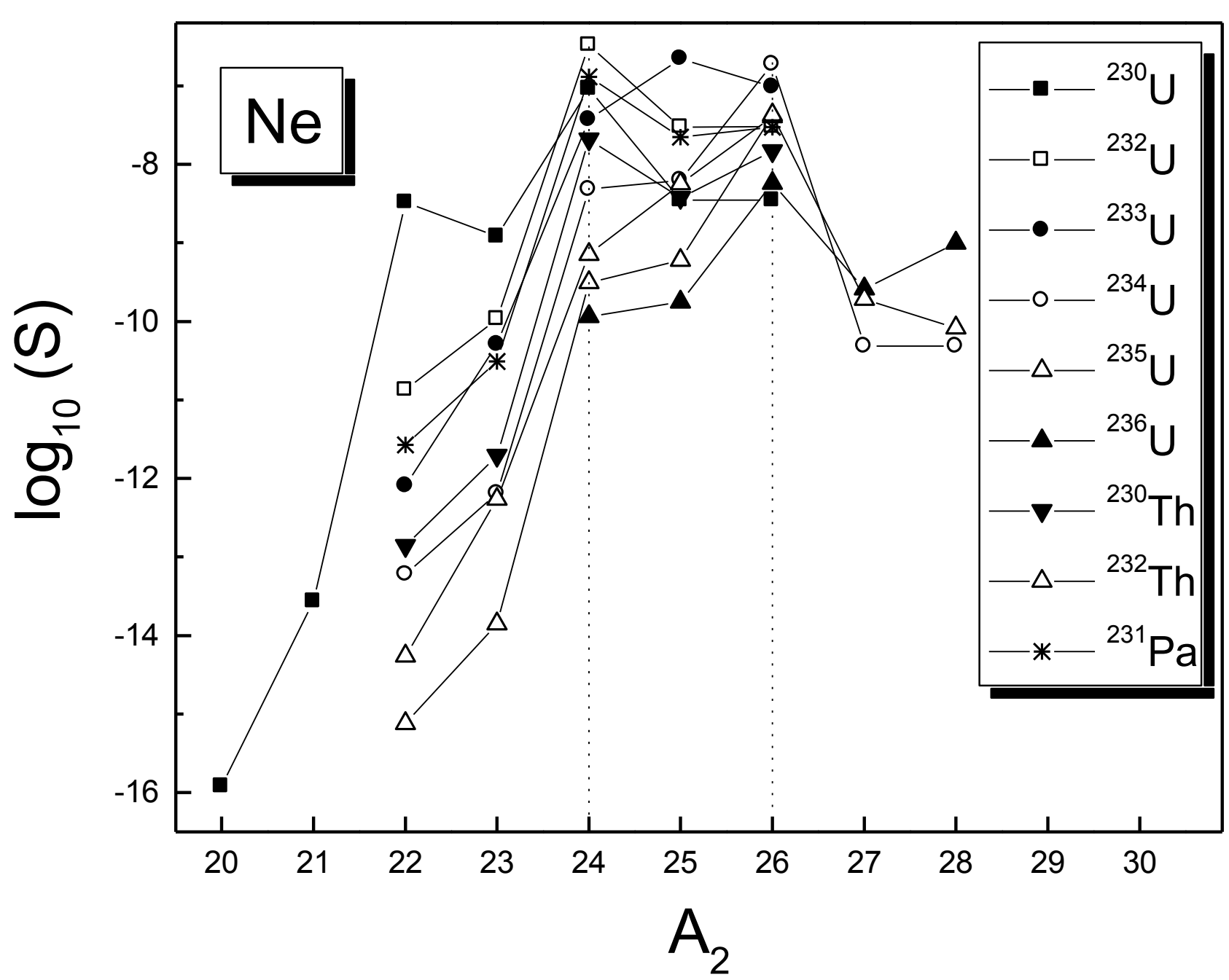

**Fig. 4.** Plot for $\log_{10}$ (S) versus $A_2$, the mass number of various neon clusters from U, Th and Pa parents.

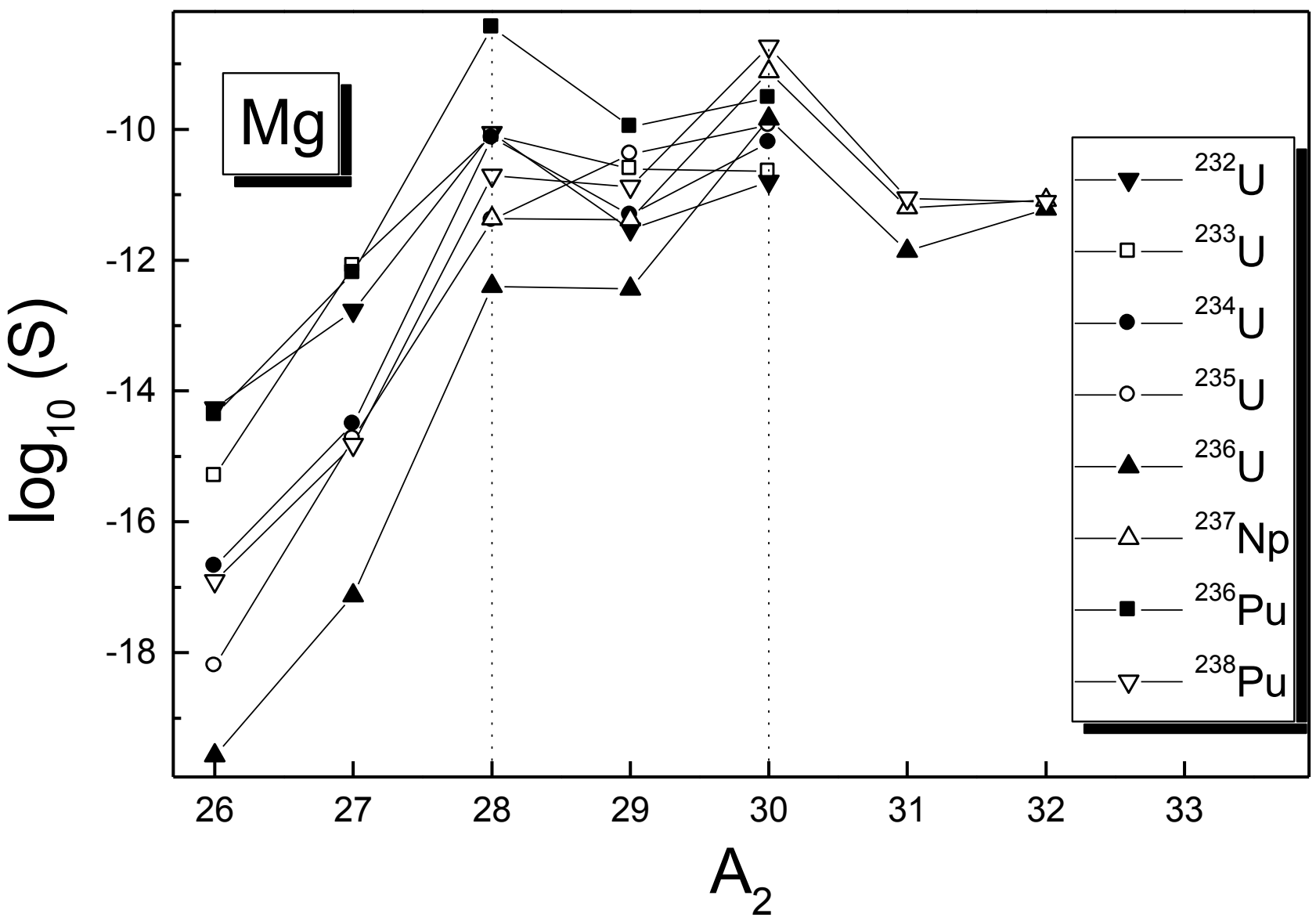

**Fig. 5.** Plot for $\log_{10}$ (S) versus $A_2$, the mass number of various magnesium clusters from U, Np, and Pu parents.

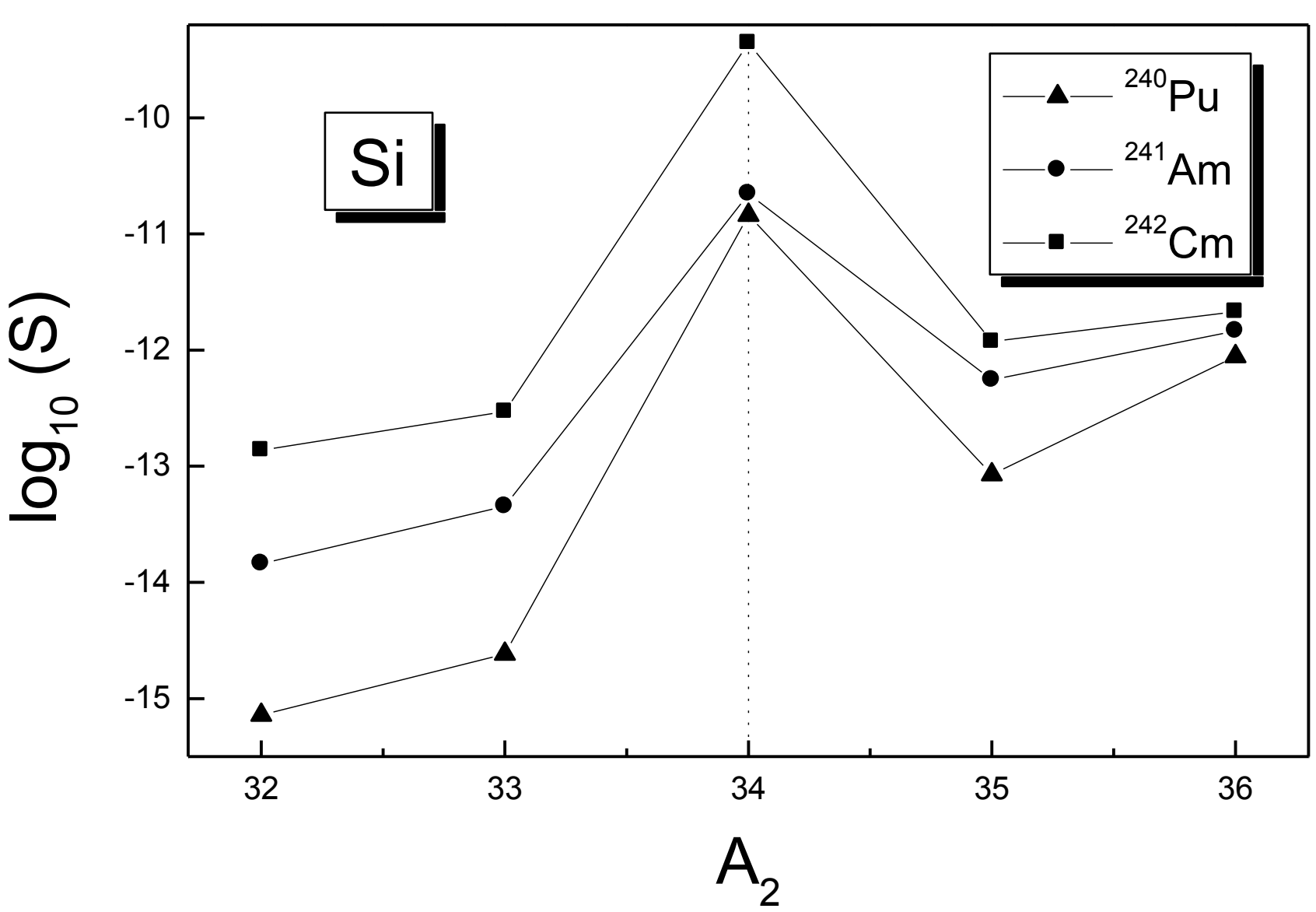

**Fig. 6.** Plot for $\log_{10}$ (S) versus $A_2$, the mass number of various silicon clusters from $^{240}$Pu, $^{241}$Am and $^{242}$Cm parents.